\documentclass[%
 reprint,
 amsmath,amssymb,
 aps,
]{revtex4-2}

\usepackage{graphicx}
\usepackage{dcolumn}
\usepackage{bm}

\begin{document}

\preprint{APS/123-QED}

\title{Unconditional Microwave Quantum Teleportation of Gaussian States in Lossy Environments}

\author{Vahid Salari}
\affiliation{%
 Department of Physical Chemistry, University of the Basque Country UPV/EHU, Apdo. 644, 48080 Bilbao, Spain
}%




\date{\today}

\begin{abstract}
Here, a physical formalism is proposed for an unconditional microwave quantum teleportation of Gaussian states via two-mode squeezed states in lossy environments. The proposed formalism is controllable to be used in both the fridge and free space in case of entanglement between two parties survives. Some possible experimental parameters are estimated for the teleportation of microwave signals with a frequency of 5GHz based on the proposed physical framework. This would be helpful for superconducting inter- and intra-fridge quantum communication as well as open-air quantum microwave communication, which can be applied to quantum local area networks (QLANs) and distributed quantum computing protocols. 
\end{abstract}

\maketitle


$Introduction-$ In recent years, several acheivements have been reported~\cite{Displacement, Goetz1, Goetz2, Pogorzalek, SanzIEEE, Federov}, which are now the potential building blocks of microwave quantum communication protocols. This novel research area is not only useful for free space communications, but also a candidate for chip-to-chip communication, required by distributed quantum computing. The latter represents an alternative paradigm to increasing the number of qubits in a single quantum processor, and aims at solving larger quantum algorithms in a distributed form between different processors with lower number of qubits. Nowadays, one of the best quantum platforms suited for quantum computing is superconducting circuits, which interact via microwave photons. The main challenge concerns the efficient distribution of such microwave states between circuits, the two main options being direct state transfer and teleportation. Concerning the former, several experiments have been done in a single cryogenic environment~\cite{Axline, Campagne, Kurpiers, Leung, Roch, Narla, Dickel}, as well as with microwave to optical conversion~\cite{Forsch, Rueda}. Recently, a successful transfer of transmon qubits has been reported, via a cryogenic waveguide coherently linking two dilution refrigerators separated by five meters, with average transfer and target state fidelities of 85.8 \% and 79.5 \%, respectively, in terms of discrete variables~\cite{SuperQLAN}. 

So far, microwave teleportation between two fridges has not been investigated experimentally, neither in terms of discrete variables nor continuous variables (CVs). However, a teleportation scheme in the microwave regime in terms of CVs has been proposed before~\cite{Roberto}. Taking into account the limitations of the previous protocol, we investigate the feasibility of an experimental implementation of microwave quantum teleportation of Gaussian states in real conditions, e.g. fridge and free space, in terms of CVs based on a clear physical formalism. Indeed, in quantum information processing with CVs, Gaussian states play important roles~\cite{Adesso1, Ferraro}, which are well-known and commonly-used experimentally. Here, we focus on teleportation of single-mode Gaussian states via entangled two-mode Gaussian states.

$Quantum\;\;Teleportation\;\;with\;\;CVs-$A preliminary CV model for teleportation was proposed by Vaidman~\cite{Vaidman}, and then developed by Braunstein and Kimble~\cite{Braunstein}. The latter represented a conditional teleportation protocol, whereas an unconditional one was proposed by Furusawa et al.~\cite{Furusawa1}. 

\begin{figure}
\includegraphics[scale=0.3]{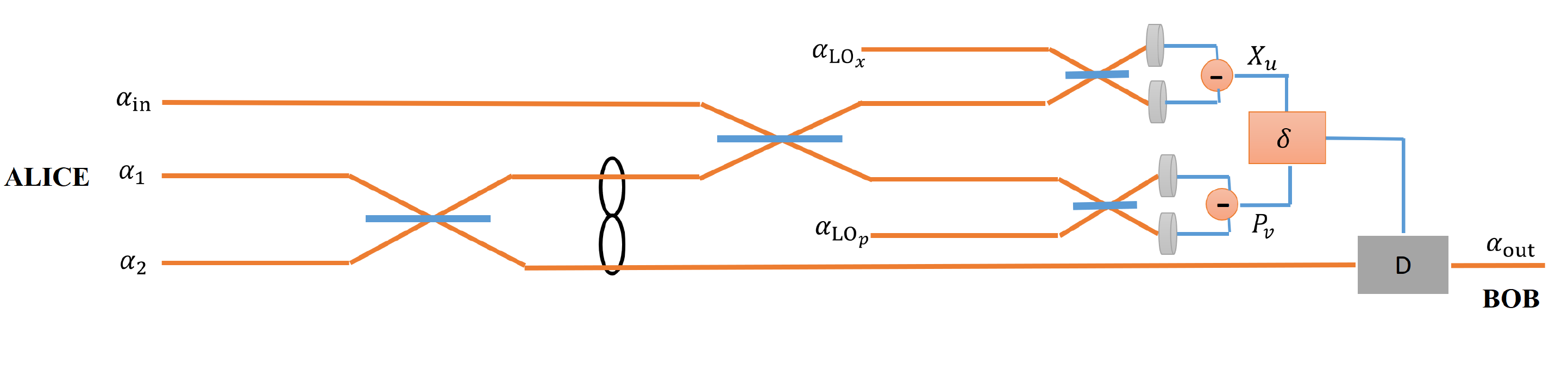}
\caption{The Braunstein-Kimble Protocol for the Teleportation of Gaussian States. In this protocol, two parties, Alice and Bob, share an entangled quantum state with two modes $\alpha_1$ and $\alpha_2$. In fact, Alice wants to send a single-mode Gaussian state (i.e. mode $\alpha_{\text{in}}$) to Bob via the teleportation mechanism where she cannot send the original mode $\alpha_{\text{in}}$ to Bob directly but a classical mode $\delta$ is sent instead. Finally, Bob can reconstruct the input state by applying the displacement $D(\delta)$ operator on his mode $\alpha_2$ to construct the input state as $\alpha_{\text{out}}\approx \alpha_{\text{in}}$.}
\label{fig1}
\end{figure}

Quantum teleportation uses quantum entanglement and classical communication to transfer quantum information between two distant parties, Alice and Bob. The ideal scenario involves Alice and Bob sharing a maximally-entangled state, which in CVs translates to a two-mode squeezed vacuum (TMSV) state with infinite squeezing $r$. In a realistic scenario, the squeezing $r$ has technological limitations. In such protocol, Alice attempts to send a Gaussian state, with covariance matrix $ V'_{\text{in}}$ and first moments $ \bar{x}^{\prime}_{\text{in}}=( x_{\text{in}}, p_{\text{in}})$ (see Appendix) to Bob. The input state can also be described by a harmonic oscillator mode $\alpha_{\text{in}}=\frac{x_{\text{in}}+ip_{\text{in}}}{\sqrt{2}}$. If this process is successful, then Alice and Bob share a Gaussian entangled state with covariance matrix ${ V_{\text{TMSS}}}$ and null first moments~\cite{Serafini}. Then, after entangling the input state with TMSS, Alice makes a double homodyne measurement of both quadratures (i.e. a heterodyne detection) and modulates the classical results, $X_{\text{u}}$ and $P_{\text{v}}$, in the form of a single mode $\delta=\frac{X_{\text{u}}+iP_{\text{v}}}{\sqrt{2}}$ and then sends it to Bob.  Finally, Bob applies a unitary displacement with a function of $\delta$ to his share of the original entangled state to reconstruct the input state (see Fig.~\ref{fig1}). Different teleportation protocols in terms of CVs are discussed in Ref.~\cite{Mancini}. In the following, we focus basically on the Braunstein-Kimble protocol~\cite{Braunstein} for teleportation of Gaussian states.

\begin{figure}
\includegraphics[scale=0.35]{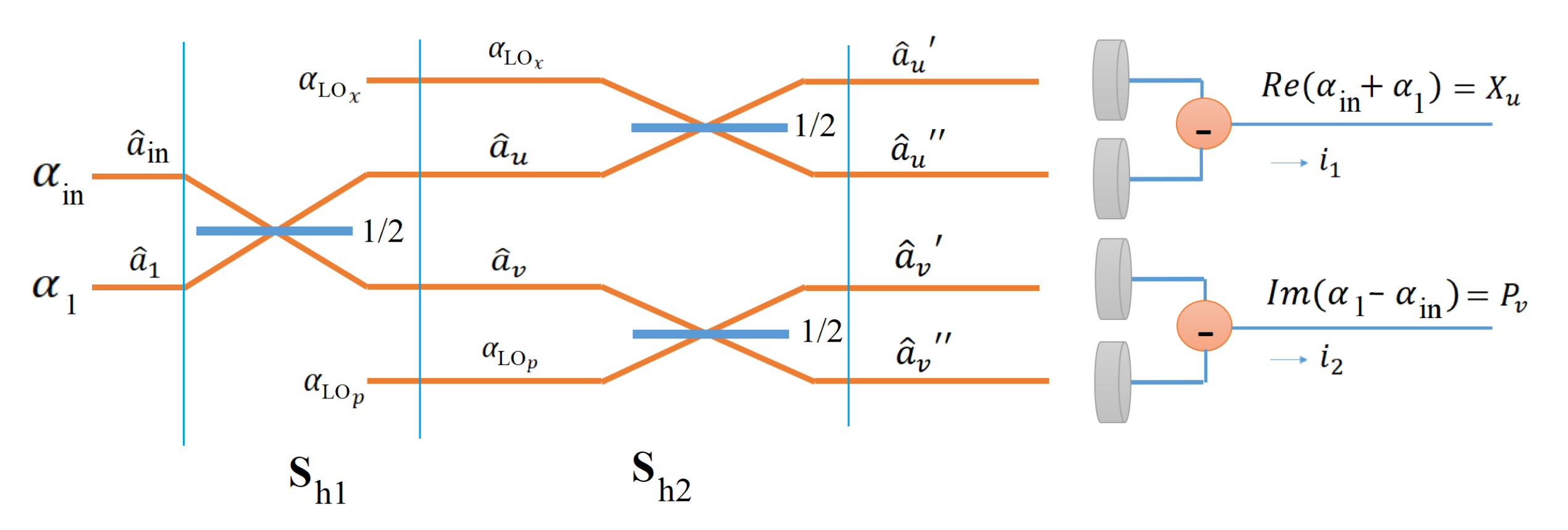}
\caption{Alice's measurement is performed via double homodyne detection. First, two modes $\alpha_{\text{in}}$ and $\alpha_1$ passing through a 50:50 beamsplitter, and then again each output beam again passes through two other 50:50 beamsplitters while a classical local oscillator mode $\alpha_{\text{LO}_x}$ enters in the top beamsplitter and another classical local oscillator mode $\alpha_{\text{LO}_p}$ passes through the other beamsplitter. Then, Alice performs a double homodyne detection where the classical result of the measurement will be modulated as a single mode $\delta$ to be sent to Bob.}
\label{fig2}
\end{figure}

$Measurement\;\;by\;\;Alice-$When Alice receives the mixture of input mode and one mode of the entangled state, she performs a double homodyne detection (see Fig.~\ref{fig2}), which is the optimal measurement for the teleportation protocol. In this case, the results of the measurement (which are two classical values, $X_u$ and $P_v$) will be modulated as a single mode $\delta$ to be sent to Bob. As seen in Fig.~\ref{fig2}, two modes $\alpha_{\text{in}}$ and $\alpha_1$ are passing through a 50:50 beamsplitter, and then again each output beam again passes through two other 50:50 beamsplitters while a classical local oscillator mode $\alpha_{\text{LO}_x}$ enters in the top beamsplitter and another classical local oscillator mode $\alpha_{\text{LO}_p}$ passes through the other beamsplitter. According to Fig.~\ref{fig2} in each line for quantum modes we use the annihiliation operators, and $\alpha$ as classical coherent state for the local oscillator. Finally, there would be four outputs at the end, where there is a detector at each output which can measure the current produced due to the collision of photons to the detector.

In the symplectic representation (see Appendix), we consider two symplectic operators in the double-homodyne detection circuit depicted in Fig.~\ref{fig2}, $ S_{\text{h1}}= \mathbb{I}_2 \oplus B_{\text{S}}(1/2)_4 \oplus \mathbb{I}_2$ and $ S_{\text{h2}}= B_{\text{S}}(1/2)_4 \oplus B_{\text{S}}(1/2)_4$ where $\mathbb{I}_2$ is the identity 2$\times$2matrix, and $B_{\text{S}}(1/2)_4$ is the 50:50 beamsplitter operator, i.e. a 4$\times$4 matrix. The input first moment in the double-homodyne detection is ${ {\bf \bar{x'}}_{\text{in}}}=( x_{\text{LO}_x}, p_{\text{LO}_x}, x_{\text{in}}, p_{\text{in}}, e^{r}x_1, e^{-r}p_1, x_{\text{LO}_p}, p_{\text{LO}_p})^{ T}$, and therefore the output's first moments at the photodetectors right before the measurement are $ { \bar{x'}}_{\text{out}}= S_{\text{h2}}S_{\text{h1}}{ \bar{x'}}_{\text{in}}$, which gives ${ { \bar{x'}}_{\text{out}}}=\sqrt{2}( {(x_{u'}), (p_{u'}), (x_{u''}), (p_{u''}), (x_{v'}), (p_{v'}), (x_{v''}), (p_{v''}))^T}$.

Here, we assume that the detectors are ideal, therefore we define the produced current as $i= \langle \hat{n}\rangle=\langle \hat{a}^\dagger\hat{a}\rangle=\langle 1/2(\hat{x}^2+\hat{p}^2)-1/2\rangle=1/2(x^2+p^2)-1/2$. Then, the difference between the currents from each detector after the beam splitters is measured~\cite{Braunstein2}. The current from the top beamsplitter is the difference between the currents in $u'$ and $u''$, i.e.    
$i_1=\langle\hat{a}^{\dagger}_{u'}\hat{a}_u'\rangle-\langle\hat{a}^{\dagger}_{u''}\hat{a}_{u''}\rangle$, 
and in the other two arms $i_2=\langle\hat{a}^{\dagger}_{v'}\hat{a}_{v'}\rangle-\langle\hat{a}^{\dagger}_{v''}\hat{a}_{v''}\rangle$.
Assuming $\alpha_{\text{LO}_x}=|\alpha_{\text{LO}_x}|e^{i\theta_x}$, one obtains $i_1=|\alpha_{\text LO_x}|(x_{\text{in}}+e^rx_1)$ by letting $\theta_x=0$ that consequently can obtain $X_u=\frac{i_1}{|\alpha_{\text LO_x}|}=x_{\text{in}}+e^rx_1=\text{Re}(\delta)$. Similarly $\alpha_{\text{LO}_p}=|\alpha_{\text{LO}_p}|e^{i\theta_p}$, if we let $\theta_p=\pi/2$ then it turns to $i_2=|\alpha_{\text LO_p}|(p_{\text{in}}-e^{-r}p_1)$, and therefore $P_v=\frac{i_2}{|\alpha_{\text LO_p}|}=p_{\text{in}}-e^{-r}p_1=\text{Im}(\delta)$. By using classical coherent light with similar amplitudes for both local oscillators, one can write $|\alpha_{\text LO_x}|=|\alpha_{\text LO_p}|=|\alpha_{\text LO}|$, so the currents $i_1$ and $i_2$ can be modulated into a single mode as $\delta=X_u+iP_v$, to be sent and received by Bob. On the other side, at the same time that Alice measures her state, the state at Bob collapses to $\alpha_2=\frac{e^rx_2+ie^{-r}p_2}{\sqrt{2}}$ with $x_2=-x_1$ and $p_2=p_1$. Comparing $\delta$ and $\alpha_2$ we realize that $\delta+\alpha_2=\alpha_{\text{in}}$. Once Bob receives this information, he can reconstruct Alice’s input state by applying the displacement $D(\delta)$ on his mode, i.e. $D(\delta) |\alpha_2\rangle=|\delta+\alpha_2\rangle=|\alpha_{\text{out}}\rangle\approx |\alpha_{\text{in}}\rangle$. In the symplectic representation, this means that the first moments of Bob, ${\bar{x}}_2=(x_2, p_2)^T$ should be displaced with ${ \Delta}=(X_u, P_v)^T$, which means
\begin{equation}
{ \bar{x}}_2 \rightarrow { \bar{x}}_2+{ \Delta}.
\end{equation}
The performance of the teleportation protocol can be measured by the teleportation fidelity $F$. It can be computed via a symplectic approach~\cite{Weedbrook, Mancini}, i.e. $F=2/\sqrt{\text{det}({ \Gamma})}$, in which $ \Gamma=2{ V'_{\text{in}}+\mathbb{Z}A\mathbb{Z}+B-C\mathbb{Z}-\mathbb{Z}^TC^T}$ where $A$, $B$, and $C$ are the block matrices of two-mode squeezed states (TMSS) in symplectic representation, i.e. shown by a block matrix as
$ V_{\text{TMSS}}=\left[A, C; C^T, B\right]$, where $ A=A^T$, $ B=B^T$ and $ C$ is a $2\times 2$ real matrix, and $T$ denotes transpose~\cite{Serafini}. If the input is a squeezed coherent state, i.e. ${V'_{\text{in}}}=\big(\begin{smallmatrix} e^{2y} & 0\\ 0 & e^{-2y} \end{smallmatrix}\big)$ where $y$ is the squeezing level of the input, the fidelity in general form is obtained as 
\begin{equation}	
F=\frac{1}{\sqrt{(e^{-2y}+(2n+1)\sigma)(e^{+2y}+(2n+1)\sigma)}}.
\end{equation}
where $\sigma=\exp{(-2r)}$ is the variance of the resource, and $n$ is the number of thermal photons in the two-mode squeezed thermal states (TMSTS) resource as a general form for TMSS. The particular case $n=0$ represents the fidelity for the general case when the resource is TMSV (see Appendix). One can compute the average fidelity for the teleportation of an arbitrary squeezed displaced vacuum state by integration over fidelity in the range 0 and 1.

\begin{figure}
\includegraphics[scale=0.32]{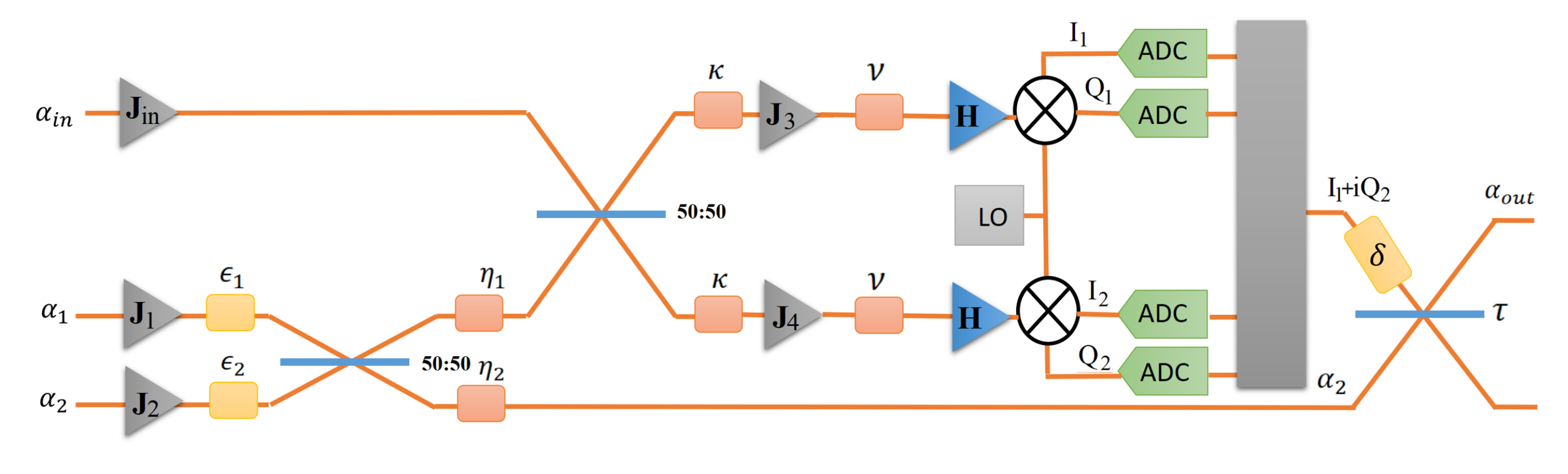}
\caption{Microwave quantum teleportation circuit for Gaussian states}
\label{fig3}
\end{figure}

$\;Microwave\;\;Quantum\;\;Teleportation-$ Microwave quantum communication, as an exciting line of research with potentially broad applications in science and industry, is an accessible technology due to the recent achievements of circuit quantum electrodynamics (cQED). Since thermal noise in the microwave domain are much larger than in the optical one, losses have to be taken into account, which can significantly affect the quality of the teleportation protocol. In order to suppress thermal fluctuations, macroscopic superconducting circuit operate at low temperatures, i.e. $T<10-100$mK~\cite{Roberto}. In cQED, superconducting Josephson junctions are nonlinear elements which have essential applications in quantum computation and quantum information. Recently, path-entanglement between propagating quantum microwaves as TMSS was generated via Josephson parametric amplifiers (JPAs) and hybrid ring \cite{Pogorzalek} to be used to perform microwave protocol equivalent to the traditional ones in optical quantum teleportation. In this protocol, a TMSS is generated from two single-mode squeezed states, with squeezing in orthogonal quadratures, generated by two JPAs, J$_1$ and J$_2$ at the same squeezing level $r$ with possible endogenous thermal photons, and then sending them through a hybrid ring, which is a microwave beam splitter~\cite{Roberto, Pogorzalek}. In general, the quality of the entanglement between the two modes is affected by thermal fluctuations on the JPA during the generation of the single-mode squeezed states. Then, after the interaction of one mode with the input state via 50:50 beamsplitter, the outputs are connected to other two JPAs, J$_3$ and J$_4$, which operate as amplifiers with equal gain $g_{\text J}=\exp(2r_{\text J})$ with squeezing parameter $r_{\text J}$ (see Fig.~\ref{fig3}). The final step is the measurement by Alice via heterodyne detection. The method is the same as what we explained before (and in Appendix) but with taking amplifications and losses into account. Transfer efficiencies are modelled by beamsplitters with reflectivities $\epsilon$, $\eta$, $\kappa$, and $\nu$, where the losses are indeed $1- \epsilon$, $1-\eta$, $1-\kappa$, and $1-\nu$, respectively. The reflectivity $\eta$ represents the interaction of TMSS in free space. Other losses are related to JPAs and amplifiers. For simplicity, we choose $\epsilon_1=\epsilon$, $\epsilon_2=1$, $\eta_1=\eta$, and $\eta_2=1$. If the single-mode operator J$_{\text {in}}$ squeezes the input coherent state with squeezing parameter $y$, e.g. ${ J}_{\text {in}}=[e^{-y}, 0; e^{y}, 0]$, it produces a squeezed coherent state as $\alpha_{in}=(e^{-y}x_{\text{in}}+ie^{y}p_{\text{in}})/\sqrt{2}$, which is the state to be teleported. Following the procedure in heterodyne detection, the output components can be obtained as $ X_u =  |\alpha_{\text{LO}}| \sqrt{\nu\kappa g_{\text {J}}} [ e^{-y} x_{\text{in}}+(e^rx_1 \sqrt{\eta\epsilon}+\zeta_{x_1})]\cos (\theta_x)$ and $P_u =  |\alpha_{\text{LO}}| \sqrt{\frac{\nu\kappa}{g_{\text {J}}}} [e^{ y} p_{\text{in}}+(e^{-r}p_1 \sqrt{\eta\epsilon}+\zeta_{p_1})] \sin (\theta _x)$
as the real and imaginary parts, respectively. By letting $\theta_x=0$ the current $i_1$ turns to 

\begin{equation}
X_u= [|\alpha_{\text{LO}}|(\sqrt{\nu\kappa g_{\text {J}}}  (e^rx_1\sqrt{\eta\epsilon}+e^{- y} x_{\text{in}})+\zeta_{x})]  
\end{equation}

and $P_u=0$ where $\zeta_{x}$ is the noise term, i.e. $\zeta_{x}=x_{\text{th-1}}\sqrt{\nu\kappa\eta(1-\epsilon)}+x_{\text{th-2}}\sqrt{\nu\kappa(1-\eta)}+x_{\text{th-3}}\sqrt{\nu(1-\kappa)}+x_{\text{th-4}}\sqrt{1-\nu}$ where $x_{\text{th-i}}$ ($i$=1, 2, 3, 4) are thermal quadratures at temperatures $T_1$, $T_2$, $T_3$, and $T_4$ respectively. If there are no losses ($\epsilon=\eta=\kappa=\nu=1$), then the current is $i_1= |\alpha_{\text{LO}}|\sqrt{g_{\text {f}}}  (e^rx_1+e^{- y} x_{\text{in}})$. Similarly for current $i_2$, one can obtain  $X_v= |\alpha_{\text{LO}}|\sqrt{\frac{\nu\kappa}{g_{\text {J}}}}  [e^{- y} x_{\text{in}}-(e^{r}x_1 \sqrt{\eta\epsilon}-\zeta_{x_2}) ] \cos (\theta_p)$ and $P_v= |\alpha_{\text{LO}}| \sqrt{\nu\kappa g_{\text {J}}} [e^{ y} p_{\text{in}}-(e^{-r}p_1 \sqrt{\eta\epsilon}-\zeta_{p_2})] \sin (\theta _p)$. Again, under similar conditions, but letting $\theta_p=\pi/2$ the current $i_2$ turns to 
\begin{equation}
P_v= |\alpha_{\text{LO}}|[\sqrt{\nu\kappa g_{\text {J}}}  ( e^{ y} p_{\text{in}}-e^{-r}p_1\sqrt{\eta\epsilon})+\zeta_{p}]
\end{equation}
and $X_v=0$, where the noise term $\zeta_{p}$ is $\zeta_{p}=p_{\text{th-1}}\sqrt{\nu\kappa\eta(1-\epsilon)}+p_{\text{th-2}}\sqrt{\nu\kappa(1-\eta)}+p_{\text{th-3}}\sqrt{\nu(1-\kappa)}+p_{\text{th-4}}\sqrt{1-\nu}$. If there is no any loss, then the noise term turns to zero, and $i_2=|\alpha_{\text{LO}}|\sqrt{g_{\text {J}}}( e^{ y} p_{\text{in}}-e^{-r}p_1)$. 
 
 The above results are in agreement with the heterodyne outputs (see the Appendix) by letting $g_{\text {J}}=1$ and $y=0$ for coherent states in the BK protocol. In reality, there are always losses for microwave signals. So, it is desirable to reduce the losses as far as possible to have a successful quantum protocol, otherwise the noisy signal received by Bob makes it harder to reconstruct the input state. In fact, the realization of a microwave single-photon detector is a difficult task due to the low energy of microwave photons, therefore measuring a quadrature of a weak microwave signal is hard. Thus, amplification of the signal is required. Cryogenic high electronic mobility transistor (HEMT) amplifiers are routinely used in quantum microwave experiments because of their large gains in a relatively broad frequency band. Basically, HEMT amplifiers are phase insensitive and add a significant amount of noise photons that may disrupt the protocol (see section E. on Amplification). These signals are digitized with analog-to-digital (ADC) converters and sent to a computer for digital data processing~\cite{Pogorzalek}. In terms of ADC in heterodyne detection, quadrature moments are $I_1$, $ I_2$, $Q_1$, and $Q_2$ (see Fig.~\ref{fig3}) which are calculated and averaged in the computer. The $I$ and $Q$ components can be described in terms of continuous variables $x$ and $p$, so they can be written as $I_i=\sqrt{\hbar \omega_iBRg_{\text H}}x_i$ and $Q_i=\sqrt{\hbar \omega_iBRg_{\text H}}p_i$, where $R=50\Omega$, $B$ is the measurement bandwidth set by a digital filter, and $g_{\text H}$ is the HEMT gain.
The modulated classical signal $\delta=I_1+iQ_2=\sqrt{\hbar \omega_iBRg_{\text H}}(X_u+iP_v)$ is communicated classically with Bob, where
\begin{eqnarray}
\nonumber I_1 &=&  \sqrt{\hbar \omega_iBRg_{\text H}}   [|\alpha_{\text{LO}}|(\sqrt{\nu\kappa g_{\text {J}}}  (e^{- y} x_{\text{in}}+\sqrt{\eta\epsilon}e^rx_1+\zeta'_{x})], \\
Q_2 &=&   \sqrt{\hbar \omega_iBRg_{\text H}} [|\alpha_{\text{LO}}|(\sqrt{\nu\kappa g_{\text {J}}}  ( e^{ y} p_{\text{in}}-\sqrt{\eta\epsilon}e^{-r}p_1+\zeta'_{p})].
\end{eqnarray}

where $\zeta'_{x}=\frac{\zeta_{x}}{\sqrt{\nu\kappa g_{\text J}}}$, and $\zeta'_{p}=\frac{\zeta_{p}}{\sqrt{\nu\kappa g_{\text J}}}$. The above states $I_1$ and $Q_2$ can be sent (or prepared at distant) to Bob via remote state preparation (RSP)\cite{Pogorzalek}. Then Bob displaces his mode, i.e. $\alpha_2=(e^rx_2+ie^{-r}p_2)/\sqrt{2}$, according to the received signal. In fact, after the measurement the values of entangled states already turn to $x_2=-x_1$ and $p_2=p_1$. The displacement on Bob’s side is implemented with a directional coupler and is described as an asymmetric beam splitter with transmissivity $\tau$, which is $B_S(\tau)_4=[\sqrt{1-\tau}\mathbb{I}_2, \sqrt{\tau}\mathbb{I}_2; -\sqrt{\tau}\mathbb{I}_2, \sqrt{1-\tau}\mathbb{I}_2]$ with $\tau=1-10^{\beta/10}$, where $\beta$ is the coupling strength expressed in decibels (dB)\cite{Pogorzalek, Roberto}.  Letting the first moments before the beam splitter as ${ x_{\tau}}=(I_1, Q_2, e^rx_2, e^{-r}p_2 )^T$ by adjusting the parameter $\tau$ as

\begin{equation}
\tau=\frac{\epsilon\eta}{2}=1-\frac{1}{|\alpha_{\text{LO}}|^2(\hbar \omega_iBR\nu\kappa g_{\text {J}}g_{\text H} )}
\end{equation}
and letting $\Lambda=|\alpha_{\text{LO}}|^2\hbar \omega_iBR\nu\kappa g_{\text {J}}g_{\text H}$ therefore $\tau=1-\frac{1}{\Lambda}$. Since the maximum value of $\tau=\frac{\epsilon\eta}{2}=1/2$ then it makes a restriction on $\Lambda$ as $1<\Lambda \leq 2$ to have a feasible protocol of teleportation.

The coupling strength turns to                                                                               

\begin{equation}
\beta=10\log\frac{1}{\Lambda}
\end{equation}

Finally, after the operation $B_S(\tau)_4{ x_{\tau}}$, the state will be reconstructed in the upper arm after the beam splitter as $\sqrt{2}\alpha_{in}=e^{-                                                                                                                                                                                                                                                                                                                                                y}x_{\text{in}}+ie^{y}p_{\text{in}}$ with noise term $\zeta=\zeta'_{x}+i\zeta'_{p}$ where the components are                             
 \begin{eqnarray}                                                                        
\nonumber e^{-y}x_{\text{in}} +\zeta'_{x}= \frac{1 }{\sqrt{\Lambda}}I_1+ \sqrt{\tau} e^rx_2  \\
e^{y}p_{\text{in}}+\zeta'_{p} = \frac{1 }{\sqrt{\Lambda}}Q_2+\sqrt{\tau} e^{-r}p_2 
\end{eqnarray}

 In a lossless protocol, $\eta=\epsilon=\kappa=\nu=1$, the noise term disappears and the state can be reconstructed perfectly, but in reality the amount of loss is significant in the microwave regime. The magnitudes of noise terms $\zeta'_{x}$ and $\zeta'_{p}$ can be obtained experimentally via calibration of the setup by letting the zero input values, i.e. $e^{-y}x_{\text{in}}=0$ and $e^{y}p_{\text{in}}=0$.

\begin{widetext}

\begin{table}
\centering
\begin{tabular}{l c c}
\hline
Parameter &Symbol &Value\\
\hline
Frequency  & $\omega_i$ & 5 GHz\\
Measurement bandwidth & $B$ & 420 kHz\\
Resistance & $R$ & 50 $\Omega$\\
HEMT gain & $g_{\text H}$  & $10^4$\\
Amplification gain of JPA & $g_{\text J}$ & $10^2$ \\
Transfer efficiency (at T$_1$=40mK) & $\epsilon$ & 0.95 \\
Transfer efficiency (at T$_2$=300K, Free space) & $\eta$ & 0.10\\
Transfer efficiency (at T$_2$=4K, Fridge) & $\eta$ & 0.90\\
Transfer efficiency (at T$_3$=4K) & $\kappa$ & 0.65\\
Transfer efficiency (at T$_4$=100mK) & $\nu$ & 0.75\\

Local oscillator mode amplitude & $|\alpha_{\text{LO}}|$ &  $10^6$V/m\\
Amplification squeezing& $r_{\text{J}}$ & 2.30\\ 
Squeezing parameter of the TMSS & $r$ &   1.32 \\
Transmissivity (for T$_2$=300K, Free space) & $\tau$ & 0.095\\
Transmissivity (for T$_2$=4K, Fridge) & $\tau$ & 0.427\\
Coupling strength (for T$_2$=300K, Free space)& $\beta$ & -0.41\\
Coupling strength (for T$_2$=4K, Fridge)& $\beta$ & -2.40\\
Coefficient (for T$_2$=300K, Free space) & $\Lambda$ & 1.10 \\
Coefficient (for T$_2$=4K, Fridge) & $\Lambda$ &  1.74\\
Noise (zero input, for T$_2$=300K, Free space)& $\zeta'_x$ & (0.954)$<I_1>$+(1.152)$<x_2>$\\
Noise (zero input, for T$_2$=300K, Free space)& $\zeta'_p$  & (0.954)$<Q_2>$+(0.082)$<p_2>$ \\
Noise (zero input, for $_2$=4K, Fridge)& $\zeta'_x$ & (0.758)$<I_1>$+(2.444)$<x_2>$\\
Noise (zero input, for T$_2$=4K, Fridge)& $\zeta'_p$  &  (0.758)$<Q_2>$+(0.174)$<p_2>$\\

\hline
\end{tabular}
\caption{Some approximations for the values in the microwave teleportation circuit}
\end{table}

\end{widetext}

\begin{figure}
\includegraphics[scale=0.3]{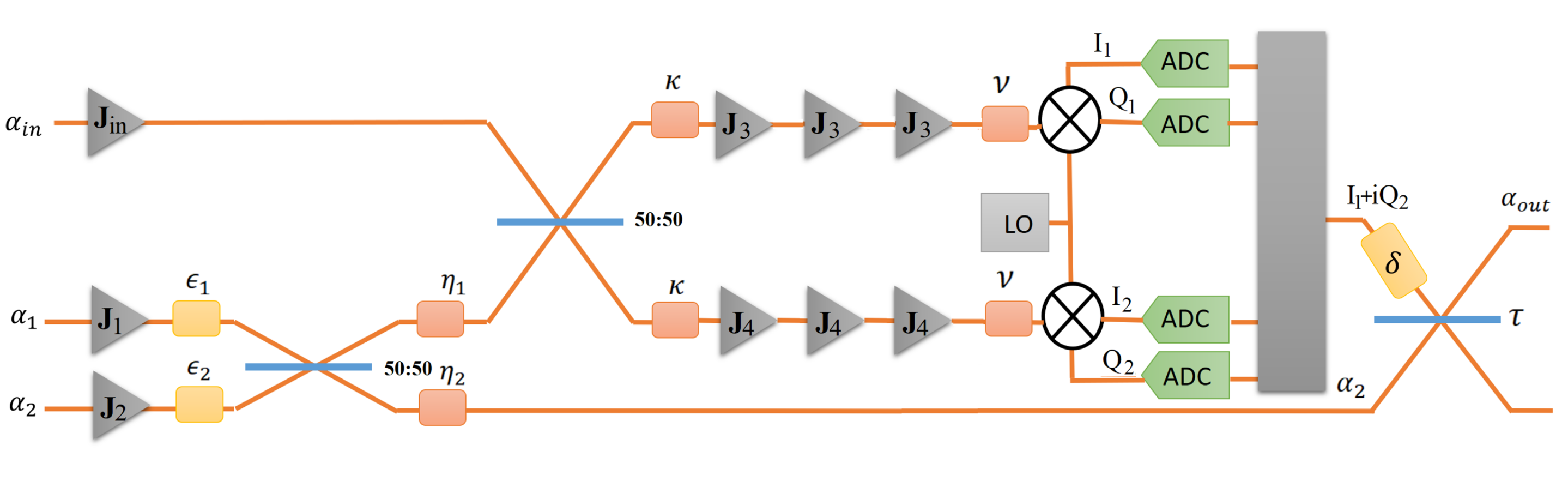}
\caption{Replacement of each HEMT with two JPAs in each arm to reach the same gain but with significant lower thermal noise.}
\label{fig4}
\end{figure}

$Amplification-$ The amplification of signals is an essential feature in microwave communication in open air. In the quantum regime, HEMT amplifiers are suited for experiments in the microwave regime. A commercial HEMT usually has $g_{H}=10^{4}$, and working at 5 GHz frequencies it introduces between $n\sim 10-100$ thermal photons~\cite{Roberto} which can have destructive effects on the protocol. One alternative for amplification is the replacement of each HMET with two additional JPAs in the same arm (see Fig.~\ref{fig4}) to reach the same gain of HEMT but with significant reduction of noise. In this case, the parameter $\Lambda$ turns to $\Lambda'$ where $\Lambda'_{\text{J}}=|\alpha_{\text{LO}}|^2\hbar \omega_iBR\nu\kappa (g_{\text {J}})^3$ and therefore the state can be reconstructed based on the circuit in Fig.~\ref{fig4}, as follows

\begin{eqnarray}                                                                        
\nonumber e^{-y}x_{\text{in}} +\zeta'_{x}= \frac{1 }{\sqrt{\Lambda'_{\text{J}}}}I_1+ \sqrt{\tau} e^rx_2  \\
e^{y}p_{\text{in}}+\zeta'_{p} = \frac{1 }{\sqrt{\Lambda'_{\text{J}}}}Q_2+\sqrt{\tau} e^{-r}p_2 
\end{eqnarray}

Based on the above theoretical proposed protocols and physical formalism, an unconditional microwave quantum teleportation is applicable in real conditions. However, quantum microwave signals are very fragile and there are experimental limitations in free space as well as some fundamental bounds \cite{SanzIEEE, Pirandolanew} that should be studied as the next research prospect.




\begin{acknowledgments}
The author greatly acknowledges financial support from the projects QMiCS (820505) of the EU Flagship on Quantum Technologies as well as EU project EPIQUS (899368). Also, the author is very grateful for very helpful and constructive discussions with Mikel Sanz, Yasser Omar, Frank Deppe, and Shabir Barzanjeh.
\end{acknowledgments}

\appendix


\section{Continuous Variables and Gaussian States}
In quantum mechanics, a continuous variable (CV) system is defined as a system with degrees of freedom associated to operators with a continuous spectrum, whose eigenstates form bases for the infinite-dimensional Hilbert space $\mathbb{H}$. For example, a quantized electromagnetic field is a bosonic CV system that can be modeled as a collection of non-interacting quantum harmonic oscillators with different frequencies, where each oscillator is referred to as a mode of the system~\cite{Adesso2, Adesso1}. CV systems are represented by $N$ bosonic modes with an infinite-dimensional Hilbert space $\mathbb{H}^{\otimes N}=\otimes_{k=1}^N\mathbb{H}_k$, corresponding to $N$ quantum harmonic oscillators associated with $N$ pairs of annihilation and creation operators $\{a_k, a_k^\dagger\}_{k=1}^N$, respectively, which can be sorted in a vectorial operator $ { \hat{l}}=(\hat{a}_1, \hat{a}^{\dagger}_1, ..., \hat{a}_N, \hat{a}^{\dagger}_N)$, satisfying in the bosonic commutation relations $[\hat{l}_i, \hat{l}_j]=\Omega_{ij}, \;\; i,j=1,..., 2N$ where $\Omega_{ij}$ is the generic element of $2N\times 2N$ matrix $ \Omega:=\oplus_{k=1}^N\omega, \;\; \omega:=\big(\begin{smallmatrix} 0 & 1\\ -1 & 0 \end{smallmatrix}\big)$ known as the symplectic form \cite{Adesso2}. Besides $\{a_k, a_k^\dagger\}$ operators, a bosonic system may be described by the quadrature field operators $\{q_k, p_k\}_{k=1}^N$, sorted in the vector $ { \hat{x}}=(\hat{q}_1, \hat{p}_1, ..., \hat{q}_N, \hat{p}_N)$, which are associated to the bosonic field operators via $\hat{a}_k:=\frac{1}{\sqrt{2}}(\hat{q}_k+i\hat{p}_k)$. The quadrature field operators act like the position and momentum operators of the quantum harmonic oscillator, satisfying in the canonical commutation relations in natural units ($\hbar=2$), $[\hat{x}_i, \hat{x}_j]=2i\Omega_{ij}$. The N-mode Hilbert space can be written as ${ \hat{x}^T}|x\rangle={ x^T}| x\rangle$ with ${ x}\in \mathbb{R}^{2N}$ and $| x\rangle:=(|x_1\rangle,...,|x_{2N}\rangle)^T$. A quantum state represented by a density operator $\hat{\rho}$ includes all the physical information about the N-mode bosonic system, which has an equivalent representation in terms of a quasi-probability distribution, i.e. Wigner function, over a real phase space. Also, it is equivalent to a Wigner characteristic function $\chi({ \xi})=\text{tr}[\hat{\rho}D({ \xi})]$ where $D({ \xi}):=\exp(i{ \hat{x}^T\Omega\xi})$ is the Weyl operator with ${ \xi}\in \mathbb{R}^{2N}$ that can be converted to a Wigner function via Fourier transform $ W(x)=\int_{\mathbb{R}^{2N}}\frac{d^{2N}\xi}{(2\pi)^{2N}}\exp(-i{ x^T\Omega\xi})\chi({ \xi})$, i.e. normalized to one but non-positive quasiprobability distribution in general~\cite{Adesso2, Weedbrook}. The CVs, ${ x}\in \mathbb{R}^{2N}$, which are the eigenvalues of quadratures operator $\hat{{ x}}$, span a real symplectic space $(\mathbb{R}^{2N}, \Omega)$, in which a N-mode bosonic state $\rho$ is equivalent to a Wigner function ${ W(x)}$ defined over a 2N-dimensional phase space. The statistical moments of the quantum state characterize $\chi$ or ${ W}$ where the first moment is called the displacement vector or the mean value ${ \bar{x}}:=\langle{ \hat{x}}\rangle=\text{tr}({ \hat{x}}\rho)$, and the second moment is called the covariance matrix ${ V}$, i.e. a $2N\times 2N$, real and symmetric matrix, with elements $V_{ij}:=\frac{1}{2}\langle\{ \Delta\hat{x}_i, \Delta\hat{x}_j\}\rangle$ where $\Delta\hat{x}_i:=\hat{x}_i-\langle\hat{x}_i\rangle$ and $\{ , \}$ is the anti-commutator. In fact, the diagonal elements of the covariance matrix are the variances of the quadrature operators, i.e., $V_{ii}=V(\hat{x}_i)$ where $V(\hat{x}_i)=\langle(\Delta\hat{x}_i)^2\rangle=\langle\hat{x}_i^2\rangle-\langle\hat{x}_i\rangle^2$. The covariance matrix must satisfy the uncertainty principle ${V}-i{ \Omega}\geq 0$, implying the positive definiteness ${ V}>0$. The first two moments are sufficient for a complete characterization, i.e. $\hat{\rho}=\hat{\rho}{ (\bar{x}, V)}$, which is used for the case of Gaussian states that are bosonic states with Gaussian Wigner representation ($\chi$ or ${ W}$). In classical physics, Gaussian functions are mostly introduced in probability theory, often under the name of ``normal distributions'', but in quantum theory, Gaussian states are very closely related to Gaussian functions, which are defined as states whose characteristic functions and quasiprobability distributions are Gaussian functions on the quantum phase space~\cite{Adesso2}. In fact, a pure state is Gaussian, if and only if, its Wigner function is non-negative, and one can obtain the Wigner function of each Gaussian state by letting its covariance matrix in the formula ${ W(x)}=\frac{\exp\{-1/2(({ x-\bar{x})^TV^{-1}(x-\bar{x})}\}}{(2\pi)^N\sqrt{\text{det}{ V}}}$\cite{Weedbrook}. 
Furthermore, a quantum operation is named Gaussian when it preserves the nature of a Gaussian state. Thus, Gaussian channels (unitaries) are those channels which preserve the Gaussian character of a quantum state. Gaussian unitaries are generated via ${ S}=\exp(-i{ \hat{H}}/2)$ from Hamiltonians ${ \hat{H}}$ which are second-order polynomials in the field operators. In terms of the quadrature operators, a Gaussian unitary is more simply described by map ${ (S, d)}: { \hat{x}}\rightarrow { {S}\hat{x}+d}$ where ${ d}\in \mathbb{R}^{2N}$ and ${ S}$ is a $2N\times 2N$ real matrix, and it is symplectic if it satisfies in ${ S\Omega S^T=\Omega}$. It can be shown that, if ${ S}$, ${ S_1}$ and ${ S_2}$ are symplectic, then ${ S^{-1}}$, ${ S^T}$ and ${ S_1S_2}$ are also symplectic~\cite{Ferraro}, with ${ S^{-1}=\Omega S^T \Omega^{-1}}$. The action of a Gaussian unitary in terms of the statistical moments, ${ \bar{x}}$ and ${ V}$, is~\cite{Ferraro, Weedbrook} 
\begin{equation}
{ \bar{x}}\rightarrow { {S}\bar{x}+d} \;\; , \; { V\rightarrow SVS^T}.
\end{equation}
In this paper, our notations are mostly in symplectic representation in terms of statistical moments.

\section{Single- and Two-mode Gaussian States}

 In this paper, we have only considered single-mode ($N=1$) and two-mode ($N=2$) Gaussian states to be used in the teleportation protocol, which will be discussed in the following sections. Now, we show the covariance matrix representation of single-mode Gaussian states, ${ V_{\text{G}}}$, as states to be processed, transferred and measured. In the following sections, $\mathbb{I}_2=\big(\begin{smallmatrix} 1 & 0\\ 0 & 1 \end{smallmatrix}\big)$ will describe the $2\times 2$ identity matrix. One can show that any $2\times 2$ symplectic matrix, as a single-mode Gaussian state with mean ${ \bar{x}}$ and covariance matrix ${ V_{\text{G}}}$, is in its most general form as ${ V}_{\text{G}}=(2n+1) R(\theta)S(2r)R(\theta)^T$~\cite{Weedbrook}. Single mode Gaussian states comprise vaccum states ($ \bar{x}=0$), ${ V_{\text{0}}}=\mathbb{I}_2$, coherent states (${ \bar{x}}\neq$0), ${ V_{\text{c}}}=\mathbb{I}_2$, thermal states ${ V_{\text{th}}}=(2n+1)\mathbb{I}_2$, squeezed vacuum states (${ {\bar{x}}}$=0) and squeezed coherent states (${ \bar{x}}\neq$0). These two have the same covariance matrices, ${ V_{\text{r}}}=\big(\begin{smallmatrix} e^{-2r} & 0\\ 0 & e^{2r} \end{smallmatrix}\big)$ and ${ V_{\text{-r}}}=\big(\begin{smallmatrix} e^{2r} & 0\\ 0 & e^{-2r} \end{smallmatrix}\big)$, where $r$ is the squeezing level, $n$ is the number of thermal photons. A rotation matrix is ${ R}(\theta)=\big(\begin{smallmatrix} \cos\theta & \;\;\sin\theta\\ -\sin\theta & \;\;\cos\theta \end{smallmatrix}\big)$, where $\theta$ is the angle of rotation and $ S(r)=\big(\begin{smallmatrix} e^{-r} & 0\\ 0 & e^{r} \end{smallmatrix}\big)$ is the squeezing matrix. Then, the covariance matrix ${ V_{\text{r}}}$ is obtained as ${ V_{\text{r}}}= S(r) S(r)^{ T}=S(2r)=\big(\begin{smallmatrix} e^{-2r} & 0\\ 0 & e^{2r} \end{smallmatrix}\big)$ that has different quadrature noise-variances, i.e., one variance is squeezed below the quantum shot-noise, while the other is anti-squeezed above it.

Gaussian states of two bosonic modes (N=2) are characterized by simple analytical formulas, which makes them the simplest states for studying properties like quantum entanglement. The covariance matrix of a two-mode Gaussian (TMG) state in symplectic representation is shown by a block matrix as follows~\cite{Serafini},
$ V_{\text{TMG}}=\left[A, C; C^T, B\right] $, where $ A=A^T$, $ B=B^T$ and $ C$ is a $2\times 2$ real matrix. Two-mode squeezed states (TMSS) are the most useful TMG states in quantum information protocols where entanglement sharing between two parties are required. The two-mode squeezing operator as a Gaussian unitary is defined as $\hat{S}_2(r)=\exp[r(\hat{a}\hat{b}-\hat{a}^\dagger\hat{b}^\dagger)]$. By applying $\hat{S}_2(r)$ to a couple of vacuum states $|0\rangle_a|0\rangle_b$, we obtain the two-mode squeezed vacuum (TMSV) state, $|\lambda\rangle=\sqrt{1-\lambda^2}\sum_{n=0}^\infty (-\lambda)^n|n\rangle_a|n\rangle_b$ where $\lambda=\tanh(r)\in[0, 1]$~\cite{Weedbrook}. In symplectic representation, TMSV state are characterized by ${ A}={ B}=\cosh(2r) \mathbb{I}_2$ and ${ C}=\sinh(2r) \mathbb{Z}$ with $\mathbb{Z}=\text{diag}(1,-1)$. The symmetric two-mode squeezed thermal state (TMST) can be expressed by ${ A}={ B}=(2n+1)\cosh(2r) \mathbb{I}_2$ and ${ C}=(2n+1)\sinh(2r) \mathbb{Z}$ with equal number of thermal photons, n, in each mode. Normally, TMSV and TMST states are the most useful Gaussian resources in quantum protocols, e.g. in quantum teleportation. 

The Wigner function of TMG states can be obtained by replacing ${ V}$ by ${ V_{\text{TMG}}}$ in the formula of ${ W(x)}$, where ${ x}$ turns to ${ x}=(x_1\;\;p_1\;\;x_2\;\;p_2)^{T}$.\\

\subsection{Measurement in Teleportation of Gaussian States}
The Braunstein-Kimble protocol describes a CV teleportation procedure with quantum states employing Gaussian resources~\cite{Braunstein}. Let us consider a situation in which two parties, Alice and Bob, want to share a TMSS. Then, Alice tries to send a single-mode Gaussian state to Bob via the teleportation mechanism. Alice performs double homodyne detection (see Fig.~\ref{fig2}), which is the optimal measurement for the teleportation protocol. In this case, the result of the measurement will be modulated as a single mode $\delta$ to be sent to Bob. As can be seen in Fig.~\ref{fig2}, two modes $\alpha_{\text{in}}$ and $\alpha_1$ are passing through a 50:50 beamsplitter, and then again each output beam again passes through two other 50:50 beamsplitters while a classical local oscillator mode $\alpha_{\text{LO}_x}$ enters in the top beamsplitter and another classical local oscillator mode $\alpha_{\text{LO}_p}$ passes through the other beamsplitter. According to Fig.~\ref{fig2} in each line for quantum modes we use the annihiliation operators, and $\alpha$ as classical coherent state for the local oscillator. Finally, there would be four outputs at the end, where there is a detector at each output which can measure the current produced due to the collision of photons to the detector.

In the symplectic representation, we consider two operators in the double-homodyne detection circuit depicted in Fig.~\ref{fig2}, ${ S}_{h1}= \mathbb{I}_2 \oplus B_S(1/2)_4 \oplus \mathbb{I}_2$ and ${ S}_{h2}= B_S(1/2)_4 \oplus B_S(1/2)_4$. The input first moment in the double-homodyne detection is ${ { \bar{x'}}_{\text{in}}}=( x_{\text{LO}_x}, p_{\text{LO}_x}, x_{\text{in}}, p_{\text{in}}, e^{r}x_1, e^{-r}p_1, x_{\text{LO}_p}, p_{\text{LO}_p})^{ T}$, and therefore the output's first moments at the photodetectors right before the measurement are ${ { \bar{x'}}_{\text{out}}}= S_{\text{h2}}S_{\text{h1}}{ \bar{x'}}_{\text{in}}$, which gives ${ { \bar{x'}}_{\text{out}}}=\sqrt{2}( (x_{u'}), (p_{u'}), (x_{u''}), (p_{u''}), (x_{v'}), (p_{v'}), (x_{v''}), (p_{v''}))^T$ where ${ {\text{Re}}(\alpha}_{u'})= x_{u'}= 1/2(x_{\text{in}}+ e^{r}x_1)+ x_{\text{LO}_x}/\sqrt{2}$, ${ {\text{Im}}(\alpha}_{u'})= p_{u'}= 1/2(p_{\text{in}}+ e^{-r}p_1)+ p_{\text{LO}_x}/\sqrt{2}$, ${ {\text{Re}}(\alpha}_{u''})= x_{u''}= 1/2(x_{\text{in}}+ e^{r}x_1)- x_{\text{LO}_x}/\sqrt{2}$, ${ {\text{Im}}(\alpha}_{u''})= p_{u'}= 1/2(p_{in}+ e^{-r}p_1)- p_{\text{LO}_x}/\sqrt{2}$, ${ {\text{Re}}(\alpha}_{v'})= x_{v'}=1/2(-x_{\text{in}}+ e^{r}x_1)+ x_{\text{LO}_x}/\sqrt{2}$, ${ {\text{Im}}(\alpha}_{v'})= p_{v'}=1/2(- p_{\text{in}}+ e^{-r}p_1)+ p_{\text{LO}_x}/\sqrt{2}$, ${ {\text{Re}}(\alpha}_{v''})= x_{v''}= 1/2(x_{\text{in}}- e^{r}x_1)+ x_{\text{LO}_x}/\sqrt{2}$, and ${ {\text{Im}}(\alpha}_{v''})= p_{v''}= 1/2(p_{\text{in}}- e^{-r}p_1)+ p_{\text{LO}_x}/\sqrt{2}$. Here, we assume that the detectors are ideal, therefore we define the produced current as $i= \langle \hat{n}\rangle=\langle \hat{a}^\dagger\hat{a}\rangle=\langle 1/2(\hat{x}^2+\hat{p}^2)-1/2\rangle=1/2(x^2+p^2)-1/2$. Then, the difference between the currents from each detector after the beam splitters is measured~\cite{Braunstein2}. The current from the top beamsplitter is the difference between the currents in $u'$ and $u''$, i.e.    
$i_1=\langle\hat{a}^{\dagger}_{u'}\hat{a}_u'\rangle-\langle\hat{a}^{\dagger}_{u''}\hat{a}_{u''}\rangle$, 
which gives $ i_1=\frac{1}{\sqrt{2}}[(x_{\text{in}}+e^rx_1)x_{{\text LO_x}}+(p_{\text{in}}+e^{-r}p_1)p_{{\text LO_x}}]$. Assuming $\alpha_{\text{LO}_x}=|\alpha_{\text{LO}_x}|e^{i\theta_x}=|\alpha_{\text{LO}_x}|(\cos\theta_x+i\sin\theta_x)$, and equivalently $\alpha_{\text{LO}_x}=(\frac{x_{{\text LO_x}}+ip_{{\text LO_x}}}{\sqrt{2}})$, if
$\theta_x=0$ then $x_{\text LO_x}=\sqrt{2}|\alpha_{\text LO_x}|$ and $p_{\text LO_x}=0$, thus $i_1=|\alpha_{\text LO_x}|(x_{\text{in}}+e^rx_1)$ that consequently one can obtain $X_u=\frac{i_1}{|\alpha_{\text LO_x}|}=x_{\text{in}}+e^rx_1=\text{Re}(\delta)$. Similarly, in the other two arms the current difference is $i_2=\langle\hat{a}^{\dagger}_{v'}\hat{a}_{v'}\rangle-\langle\hat{a}^{\dagger}_{v''}\hat{a}_{v''}\rangle$ yields $ i_2=\frac{1}{\sqrt{2}}[(x_{\text{in}}-e^rx_1)x_{{\text LO_x}}+(p_{\text{in}}-e^{-r}p_1)p_{{\text LO_x}}]$. Assuming $\alpha_{\text{LO}_p}=|\alpha_{\text{LO}_p}|e^{i\theta_p}$, if we let $\theta_p=\pi/2$ then it turns to $i_2=|\alpha_{\text LO_p}|(p_{\text{in}}-e^{-r}p_1)$, and therefore $P_v=\frac{i_2}{|\alpha_{\text LO_p}|}=p_{\text{in}}-e^{-r}p_1=\text{Im}(\delta)$. By using classical coherent light with similar amplitudes for both local oscillators, one can write $|\alpha_{\text LO_x}|=|\alpha_{\text LO_p}|=|\alpha_{\text LO}|$, so the currents $i_1$ and $i_2$ can be modulated into a single mode as $\delta=X_u+iP_v$, to be sent and received by Bob (see Fig.~\ref{fig2}). On the other side, at the same time that Alice measures her state, the state at Bob collapses to $\alpha_2=\frac{e^rx_2+ie^{-r}p_2}{\sqrt{2}}$ with $x_2=-x_1$ and $p_2=p_1$. The goal of teleportation is to reconstruct $\alpha_{\text{in}}=\frac{x_{\text{in}}+p_{\text{in}}}{\sqrt{2}}$ from the received state $\delta=\frac{(x_{\text{in}}+e^rx_1)+i(p_{\text{in}}-e^{-r}p_1)}{\sqrt{2}}$ and $\alpha_2=\frac{-e^rx_1+ie^{-r}p_1}{\sqrt{2}}$. Comparing $\delta$ and $\alpha_2$ we realize that $\delta+\alpha_2=\alpha_{\text{in}}$ where $\delta$ is the complex conjugate of $\delta$. Once Bob receives this information, he can reconstruct Alice’s input state by applying the displacement $D(\delta)$ on his mode, i.e. $D(\delta)|\alpha_2\rangle=|\delta+\alpha_2\rangle=|\alpha_{\text{out}}\rangle\approx |\alpha_{\text{in}}\rangle$. In the symplectic representation, this means that the first moments of Bob, ${ \bar{x}}_2=(x_2, p_2)^T$ should be displaced with ${\Delta}=(X_u, P_v)^T$, which means
\begin{equation}
{ \bar{x}}_2 \rightarrow { \bar{x}}_2+{ \Delta}.
\end{equation}

\subsection{Fidelity of Teleportation in Free Space}
In order to obtain the fidelity in free space, we consider that the resource is a TMSV state interacting in free space with a thermal bath characterized by $N$ thermal photons. This mechanism is modeled via a beamsplitter with reflectivity $\eta$, i.e. $ B_S(\eta)$.  In Fig.~\ref{}a in the the circuit ${\bf V'_{\text{out}}}= S'_1V''_{\text{in}}S'^T_1$ where the input is ${ V''_{\text{in}}}= (2N+1)\mathbb{I}_2 \oplus { V_{\text{TMSV}}} $ with $N$ as the number of environmental thermal photons, and ${\bf S'_1}=  B_S(\eta)_4 \oplus \mathbb{I}_2$, thus the output covariance matrix in block form is ${ V'_{\text{out}}}=[{ A'}, { C'}; { C'}, { B'} ] $, where ${ A'}=[(2N+1)(1-\eta)+\eta \cosh (2r)]\mathbb{I}_2$, ${ B'}=\cosh (2r)\mathbb{I}_2$, and 
 ${ C'}=\sqrt{\eta}\sinh (2r)\mathbb{Z}$. This interaction between a TMSV state with squeezing level $r$ and a thermal bath with $N$ thermal photons in free space can be modeled by a two-mode squeezed thermal (TMST) state with $n$ thermal photons in the resource and squeezing level $s'$. In this case, we use the local operations $L_1$ and $ L_2$ in the circuit (see Fig.~\ref{}) to obtain the parameters of (TMSV state + Air) in terms of TMST state. Then, we obtain ${ V''_{\text{out}}}= S'_2V_{\text{TMST}}S'^T_2$, where ${ S'_2}= L_1 \oplus L_2$
 

\begin{figure}[h!]
\centering
\includegraphics[width=3.2in]{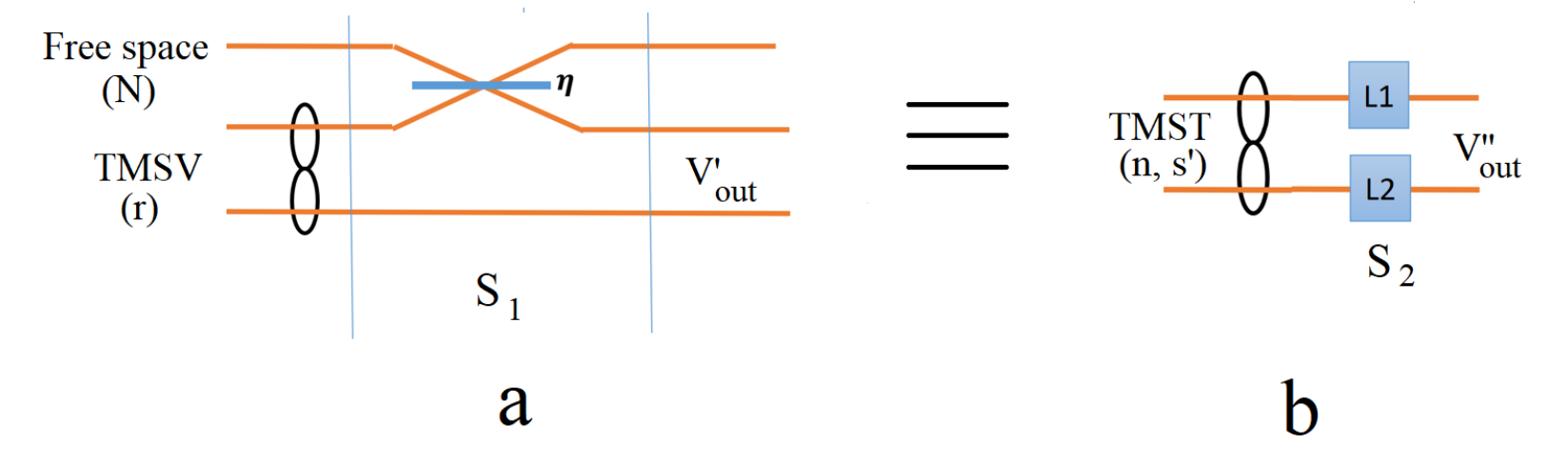}
\caption{a) The interaction model of TMSV state with squeezing level $r$ with free space including $N$ thermal photons via a beamsplitter with reflectivity $\eta$, and b) The operation of local operations ${ L_1}$ and ${ L_2}$ on the TMST state with squeezing level $s'$}
\label{}
\end{figure}
\begin{figure}[h!]
\centering
\includegraphics[width=3.5in]{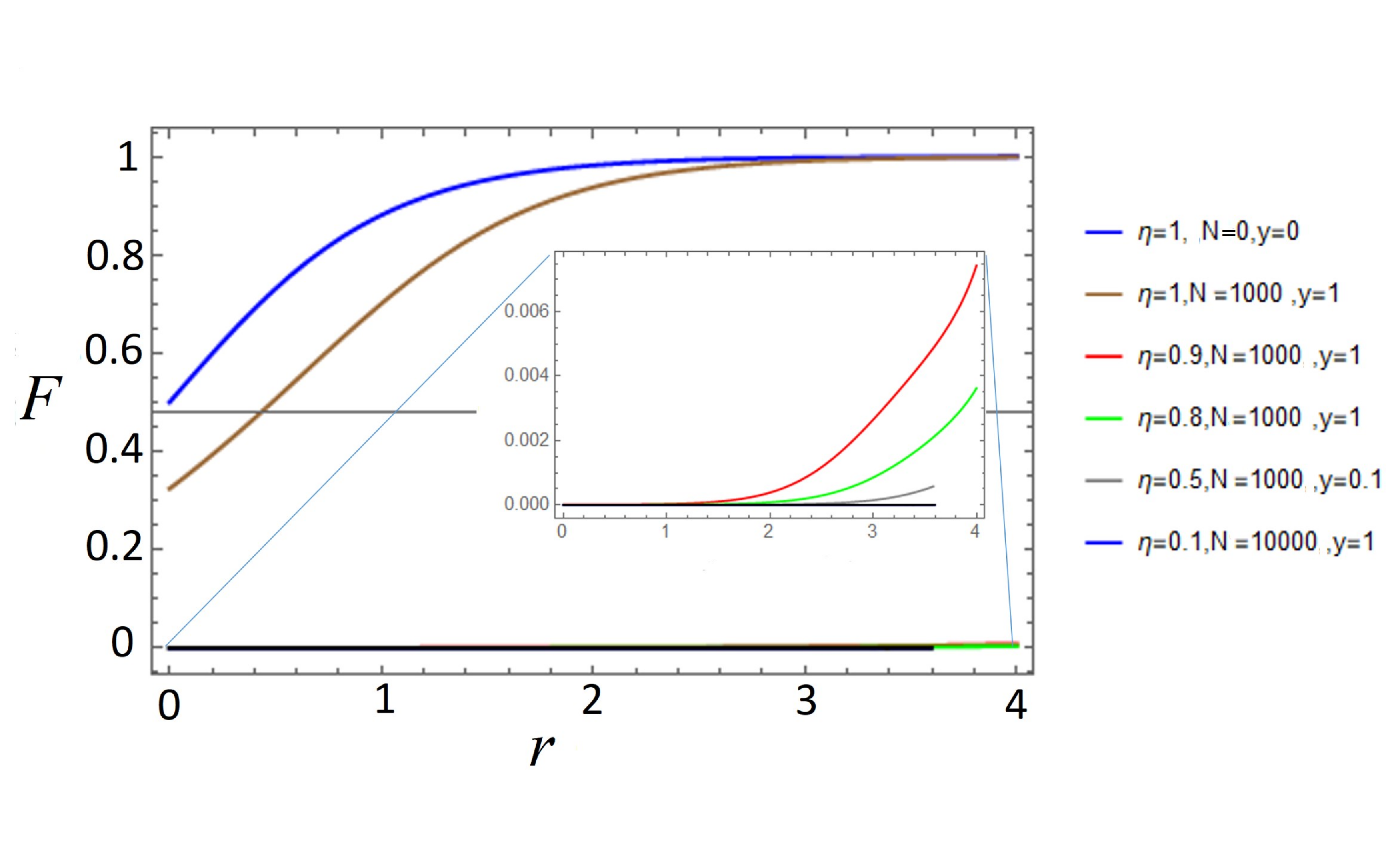}
\caption{Teleportation fidelity of Gaussian states in free space in terms of TMSV squeezing level $r$ for different values of air thermal photons $N$ and beamsplitter reflectivity $\eta$ (i.e. efficiency of transfer) for different squeezing level of the input $y$.  }
\label{fig5}
\end{figure}

After the local operations (which are squeezing operators, $L_1$ with squeezing $x_1$ and $ L_2$ with squeezing $x_2$) we have ${ V''_{\text{out}}}=[
{ A''}, {C''}; {C''}, { B''}] $, where the matrices ${ A''}$, ${ B''}$, and ${ C''}$ are 


\begin{eqnarray}
\nonumber {A''} &=& [e^{-4 x_1} (2 n+1) \cosh (2 s'), 0; 0, e^{4 x_1} (2 n+1) \cosh (2 s')], \\  
\nonumber {B''} &=& [e^{-4 x_2} (2 n+1) \cosh (2 s'), 0; 0, e^{4 x_2} (2 n+1) \cosh (2 s')], \\
{ C''} &=& [e^{-2 (x_1+x_2)} (2 n+1) \sinh (2 s'), 0; 0, -e^{2 (x_1+x_2)} (2 n+1) \sinh (2 s')]. 
\end{eqnarray}


If we equal ${V'_{\text{out}}}$ to ${V''_{\text{out}}}$, we find 

\begin{widetext}

\begin{eqnarray}
s' &=& \frac{1}{2} \tanh ^{-1}\left(\frac{\sqrt{\eta } \sinh (2 r)}{(1-\eta ) (2 \text{N}+1)+\eta  \cosh (2 r)}\right), \\ 
\nonumber x_1 &=& \frac{1}{8} \log (\text{sech}(2 r) (-\eta -2 \eta  \text{N}+2 \text{N}+\eta  \cosh (2 r)+1)), \\ 
\nonumber x_2 &=& -\frac{3}{8} \log (\text{sech}(2 r) (-\eta -2 \eta  \text{N}+2 \text{N}+\eta  \cosh (2 r)+1)), \\
\nonumber n &=& \frac{1}{2} \left(\sqrt{\text{sech}(2 r) ((\eta -1) (-2 \text{N}-1)+\eta  \cosh (2 r)) \left(((\eta -1) (2 \text{N}+1)-\eta \cosh (2 r))^2-\eta  \sinh ^2(2 r)\right)}-1\right).
\end{eqnarray} 

\end{widetext}

The term under the radical makes a restriction on the squeezing level which should satisfy
\begin{equation}
1-\frac{\eta  \sinh ^2(2 r)}{((\eta -1) (2 \text{N}+1)-\eta  \cosh (2 r))^2}>0. 
\end{equation}

The fidelity is obtained via $F=2/\sqrt{\text{det}({ \Gamma})}$, in which $ \Gamma=2{ V'_{\text{in}}+\mathbb{Z}A''\mathbb{Z}+B''-C''\mathbb{Z}-\mathbb{Z}^TC''^T}$ where $A''$, $B''$, and $C''$ are the block matrices of
$ V''_{\text{out}}=\left[A'', C''; C''^T, B''\right]$.

The diagram of the fidelity in terms of $r$, for different values of $\eta$ and $N$, and for different squeezing level $y$ of input states has been shown in Fig.~\ref{fig5}. It can be seen that the fidelity is affected significantly by variations in $\eta$ and $N$. To be quantum teleportation, the fidelity should be higher than 0.50, otherwise the protocol is classical.


\end{document}